# chi2TEX

## Полуавтоматический перевод рукописи книги из формата chiwriter в формат LaTeX

Божевольнов Юстислав
кафедра математики физического факультета МГУ

# Ключевые слова

Распознавание математических формул, chiwriter, LaTeX

# Аннотация

Современным стандартом верстки текстов научно-технического содержания является издательская система LaTeX. Этот стандарт имеет ряд достоинств, среди них: признание научным сообществом и авторитетными издательствами, высокое качество получаемого результата, поддержка множества языков и относительная простота верстки[1]. Самые популярные реализации MiKTeX[2] (windows), TeXlive[3] (linux), MacTeX[4] (MacOS) распространяются бесплатно. Формат документов LaTeXявляется *макро-форматом*: для математических формул используется набор специальных текстовых команд, охватывающих (за редким исключением) весь спектр принятых в математике структур.

Тем не менее множество рукописей по различным причинам набраны в других форматах. В частности автор столкнулся с задачей по *переводу рукописи из формата программы ChiWriter*, поддержка которой прекращена в 1996 году. Содержание рукописи (книга по физической дисциплине) не оставило выбора в конечном формате преобразования — LaTeX. Основная трудность заключалась в том, что не удалось найти сколько-нибудь подходящее средство для автоматической трансляции. Задача усложнялась отсутствием описания исходного формата.

Был произведен анализ формата документов ChiWriter. В результате "археологических раскопок" удалось получить графическую визуализацию текста, после чего стало ясно задача близка в широком смысле к задаче распознавания образов. Поскольку математические формулы весьма чувствительны к ошибкам в написании, главным стал вопрос о безошибочном разборе математических нотаций. После статистических исследований стало ясно, что *"простые" структуры можно разбирать автоматически*, но все же незначительное количество "сложных мест" требует ручного подхода. В такой постановке задачу удалось решить достаточно эффективно.

---

[1] Википедия использует именно этот макро-формат для математических формул http://en.wikipedia.org/wiki/Help:Displaying_a_formula

[2] http://miktex.org/

[3] http://www.tug.org/texlive/

[4] http://www.tug.org/mactex/

# Краткий отчет

## Исходный формат

chiwriter 3.14, WYSIWYG, прекращена поддержка в 1996 году
<http://en.wikipedia.org/wiki/ChiWriter>

## Конечный формат

LaTeX, формат де-факто для текстов научно-технического содержания, макро-формат
<http://en.wikipedia.org/wiki/LaTeX>

## Постановка задачи

- Анализ текста и, прежде всего, математических формул на возможность автоматического распознавания.

- Выработка четкого критерия для выбора
    - либо автоматического,
    - либо ручного режима распознавания.

- Безошибочное распознавание математических формул, пригодных для автоматического режима.

## Этапы работы

- *Анализ формата данных (разметка текста, шрифты) chiwriter 3.14* (реализованы классы для визуализации данных, Delphi)

- *Анализ структуры данных (текст с формулами)* (реализован программный комплекс для всевозможных статистических исследований структуры данных, Delphi)

- *Формулировка критерия выбора режима распознавания для базового элемента* (на основе проведенного статистического анализа, выработан критерий, "отсеивающий" 2% элементов для ручного распознавания)

- *Автоматический режим распознавания* (реализован конечный автомат для преобразования "псевдографических" данных в макро-формат системы LaTeX, Delphi)

- *Ручное распознавание и сведение результатов* (Delphi, Perl)

- *Приведение полученного текста к стандартам верстки в системе LaTeX* (учет стилистических особенностей верстки на русском языке, учет особенностей набора в системе LaTeX, Perl, Perl Regular Expressions)

## Статистика

Базовым элементом для преобразования взята строка в формате chi
Всего строк: около 11 000
Всего строк для разбора в ручном режиме: около 200 (2%)
Объем получившейся рукописи: около 250 страниц A4

## Результат

Рукопись в формате LaTeXпередана научному редактору.

## Иллюстрации

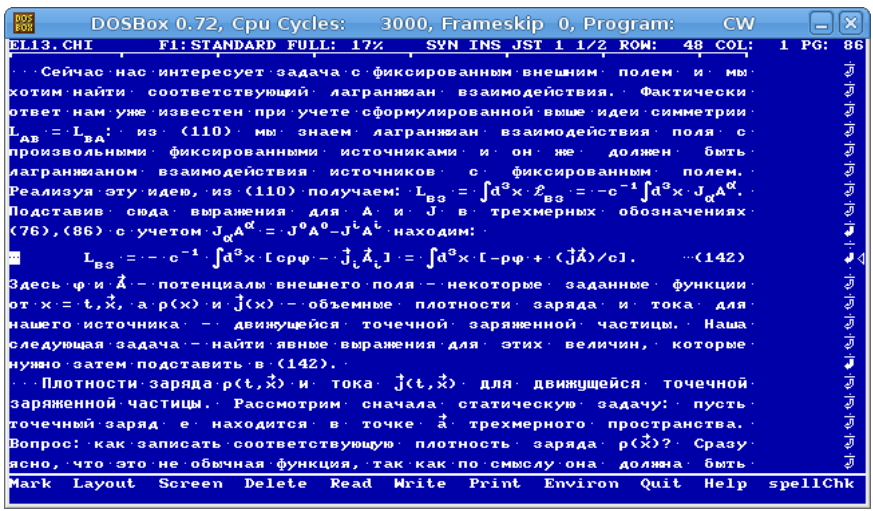

Рис. 1: Окно программы chiwriter. Если бы удалось найти матричный принтер, то примерно так бы выглядел печатный текст.

```
\+
kfuhfy;bfyjv \ dpfbvjltqcndbz \ bcnjxybrjd \ \ c \ \ abrcbhjdfyysv \ \ gjktv&
\-
\+                                 \0i \23         -1\0i \23      \7a
\5Htfkbpez 'ne blt.^ bp \1(110) \5gjkexftv\1: L\ \  = \ d\ x \#L\ \  \1= \5-\1c\ \ \ d\ x J\ A\ \5&
\-                                 dp   \0j      \5dp      \0j      \7a
\+
\5Gjlcnfdbd \ c.lf \ dshf;tybz \ lkz \ \1A \ \5b \ \1J \ \5d \ nht[vthys[ \ j,jpyfxtybz[
\-
\+                   \7a    \20 0  i i
\1(76),(86) \5c extnjv \1J\ A\  = J\ A\ -J\ A\  \5yf[jlbv\1: \,
\-                    \7a
\+
\+            \2-1 \0i \23        \9L L    \0i \23        \9LL
\^\ \ \ \ \ \ \1L\ \  = \5- \1c\ \  \ d\ x [c\7rv \1- j\ A\ ] = \ d\ x [-\7rv \1+ (jA)/c].\ \ \ \ \^(142)\,
\-       \5dp         \0j        \2i i    \0j
\-
\+       \9L
\5Pltcm \7v \5b \1A \5- gjntywbfks dytiytuj gjkz - ytrjnjhst \ pflfyyst \ aeyrwbb
\-
\+       \9L       L
\5jn \1x = t,x\5^ f \7r\1(x) \5b \1j(x) \5- j,]tvyst \ gkjnyjcnb \ pfhzlf \ b \ njrf \ lkz
\-
```

Рис. 2: "Внутренности" формата chiwriter. На примере этого фрагмента можно проследить дальнейшие метаморфозы.

Рис. 3: Возможно ли машинное распознавание? Этот вопрос потребовал более наглядного представления информации, чем на предыдущем рисунке. Для удобства исследования "на глаз" для каждого типа символов используется свой цвет.

```
Реализуя эту идею, из (110) получаем:
 $L_{\text{вз}}
 = \int d^{3} x \mathcal{L}_{\text{вз}}
 = -c^{-1} \int d^{3} x J_{\alpha } A^{\alpha }$.
Подставив сюда выражения для $A$ и $J$ в трехмерных обозначениях
(76), (86) с учетом $J_{\alpha } A^{\alpha } = J^{0} A^{0} -J^{i} A^{i}$
находим:
%%
\begin{equation*}\label{142}
   L_\text{вз}
 = - c^{-1} \int d^{3} x [c\rho \varphi - \vec{j}_{i} \vec{A}_{i} ]
 = \int d^{3} x [-\rho \varphi + (\vec{j}\vec{A})/c].
\tag{142}
\end{equation*}
%%
Здесь $\varphi$ и $\vec{A}$ --- потенциалы внешнего поля
 --- некоторые заданные функции от $x = t, \vec{x}$,
 а $\rho (x)$ и $\vec{j}(x)$ --- объемные плотности заряда и тока для
нашего источника --- движущейся точечной заряженной частицы. Наша
```

Рис. 4: Фрагмент в формате LaTeX получен в автоматическом режиме.

Сейчас нас интересует задача с фиксированным внешним полем и мы хотим найти соответствующий лагранжиан взаимодействия. Фактически ответ нам уже известен при учете сформулированной выше идеи симметрии $L_{AB} = L_{BA}$: из (110) мы знаем лагранжиан взаимодействия поля с произвольными фиксированными источниками и он же должен быть лагранжианом взаимодействия источников с фиксированным полем. Реализуя эту идею, из (110) получаем: $L_{\text{вз}} = \int d^3 x \mathcal{L}_{\text{вз}} = -c^{-1} \int d^3 x J_\alpha A^\alpha$. Подставив сюда выражения для $A$ и $J$ в трехмерных обозначениях (76), (86) с учетом $J_\alpha A^\alpha = J^0 A^0 - J^i A^i$ находим:

$$L_{\text{вз}} = -c^{-1} \int d^3 x [c\rho\varphi - \vec{j}_i \vec{A}_i] = \int d^3 x [-\rho\varphi + (\vec{j}\vec{A})/c]. \qquad (142)$$

Здесь $\varphi$ и $\vec{A}$ — потенциалы внешнего поля — некоторые заданные функции от $x = t, \vec{x}$, а $\rho(x)$ и $\vec{j}(x)$ — объемные плотности заряда и тока для нашего источника — движущейся точечной заряженной частицы. Наша следующая задача — найти явные выражения для этих величин, которые нужно затем подставить в (142).

**Плотности заряда $\rho(t, \vec{x})$ и тока $\vec{j}(t, \vec{x})$ для движущейся точечной заряженной частицы.** Рассмотрим сначала статическую задачу:

Рис. 5: Типографские каноны (печатный вариант), LaTeX

Рис. 6: Автоматический разбор строки.

Рис. 7: Слишком громоздкие конструкции пришлось "обработать напильником" в ручном режиме.

Рис. 8: Кусочек таблицы символов chiwriter. Видно, что некоторые из их составные.